\documentclass[12pt]{article}

\usepackage{amsmath}
\usepackage{amsfonts}

\newcommand{\Asla}{\ensuremath{\ A\hspace{-0.6em}/}}

\newcommand{\Psl}{\ensuremath{\ P\hspace{-0.7em}/}}
\newcommand{\psl}{\ensuremath{\ p\hspace{-0.5em}/}}
\newcommand{\bfB}{\ensuremath{\mathbf{B} }}

\newcommand{\bfr}{\ensuremath{\mathbf{r}}}
\newcommand{\bfrs}{\ensuremath{\mathbf{r}}\mbox{ }}
\newcommand{\R}{\ensuremath{\mathbb{R}}}
\newcommand{\C}{\ensuremath{\mathbb{C}}}

\numberwithin{equation}{section}

\begin{document}

\title{\textbf{QED in inhomogeneous magnetic fields}}

\author{M. P. Fry\thanks{Electronic address: mpfry@maths.tcd.ie} \\
\textit{School of Mathematics, Trinity College, Dublin 2,
Ireland }}
\date{March 1996}

\maketitle

\begin{abstract}
A lower bound is placed on the fermionic determinant of 
Euclidean quantum electrodynamics in three dimensions in the
presence of a smooth, finite--flux, static, unidirectional 
magnetic field $\mathbf{B}(\mathbf{r})=(0,0,B(\mathbf{r}))$,
where $B(\mathbf{r})\geq 0$ or $B(\mathbf{r})\leq 0$ and 
$\mathbf r$ is a point in the $xy\mbox{-plane}$.  Bounds 
are also obtained for the induced spin for $2+1$ dimensional
QED in the presence of $\mathbf{B}(\mathbf{r})$.  An upper 
bound is placed on the fermionic determinant of Euclidean 
QED in four dimensions in the presence of a strong, static,
directionally-varying, square-integrable magnetic field
$\mathbf{B}(\mathbf{r})$ on $\R^3 $. 
\\[0.2in]
PACS number(s): 12.20.Ds, 11.10.Kk, 11.15.Tk
\end{abstract}

% +++

\section{INTRODUCTION}

In quantum electrodynamics and indeed in all gauge theories coupled to
fermions the fermionic determinant is fundamental.  Without  substantially
more knowledge of this determinant a nonperturbative analysis of QED in 
the continuum with dynamical fermions will remain impossible.  The reader
is reminded that the fermionic determinant results from the integration over
the fermionic degrees of freedom in the presence of a potential $A_\mu$.
This determinant combines with the potential's gauge-fixed Gaussian measure
$d\mu(A)$ to produce a one-loop effective action $S_{eff} \propto \ln \det$
that is exact and on which every physical process in QED depends, thereby
justifying our opening statement.

In order to make this paper reasonably self-contained we will retrace 
some material previously covered in [1-4].  Schwinger's proper time
definition of the fermionic determinant in Wick-rotated Euclidean
quantum electrodynamics in four dimensions [5-7] is the most useful one
for our purpose here:
\begin{eqnarray}
\ln \mbox{det}_{ren} (1-S\hspace{-0.5em}\Asla) & = & \frac{1}{2}\int_0^{\infty} \frac{dt}{t} 
\left\{\vphantom{\frac{1}{2}} \mbox{Tr}
\left(\vphantom{\frac{1}{2}} e^{ - P^2t} \right. \right.  \nonumber \\ 
& & \left. \left. -\exp \left[ - ( D^2 +\frac{1}{2} \sigma^{\mu\nu}F_{\mu\nu} )
t \right] \right) + \frac{\|F\|^2}{24\pi^2} \right \} e^{-tm^2},
\end{eqnarray}
where $D_{\mu}=P_{\mu}-A_{\mu};$ $S$ denotes the free fermion Euclidean
propagator; $m$ is the unrenormalized fermion mass; $\sigma^{\mu\nu}
=(1/2i)[\gamma^{\mu},\gamma^{\nu}], \gamma^{\mu\dagger}=-\gamma^{\mu},$
and  $\| {F} \|^2=\int d^4x F_{\mu\nu}^2(x),$ $F_{ \mu\nu}$ being the field
strength tensor.  The coupling $e$ has been absorbed into the potential.
For future reference note that $eF_{\mu\nu}$ has the invariant dimension
of $m^2$ in any spacetime dimension.  Included in the definition is the
second-order charge renormalization subtraction at zero momentum transfer
that is required for the integral to converge for small $t$, as indicated
by the determinant's subscript.  The determinant is gauge invariant,
depending only on invariant combinations of $F_{\mu\nu}$ and their derivatives.
Definition (1.1) continues to hold for Euclidean $\mbox{QED}_3$ and 
$\mbox{QED}_2$ except
that the charge renormalization subtraction is omitted.

Now the determinant is part of a functional integral over $A_{\mu},$ and if
the gauge field is given an infrared
cutoff---a mass term---then $A_{\mu}$ is concentrated  on $\mathcal{S}',$
the Schwartz space of real-valued tempered distributions.  As we have noted
[1, 3, 4], there is a need to regulate in any dimension.  One possibility is
to replace $A_{\mu}$ in the determinant and anywhere else it appears in the 
functional integral, except in $d\mu(A),$ with the smoothed, polynomial 
bounded $C^{\infty}$ potential $A_{\mu}^{\Lambda}(x)=(h_{\Lambda} * A_{\mu})(x),$
where $A_{\mu}$ is convoluted with an ultraviolet cutoff function $h_{\Lambda}
\in \mathcal{S},$ the functions of rapid decrease [8].  This introduces a regulated photon propagator since
\begin{eqnarray}
\int d\mu(A) A_\mu^\Lambda (x) A_\nu^\Lambda (y) = D_{\mu\nu}^\Lambda(x-y),
\end{eqnarray}
where $D^\Lambda_{\mu\nu} \mbox{'s}$  Fourier transform is such that
$\hat{D}^\Lambda_{\mu\nu} \propto | \hat{h}_\Lambda |^2,$ $\hat{h}_\Lambda$ 
denoting the Fourier
transform of $h_\Lambda$.  For example, let $\hat{h}_\Lambda \in C^\infty_0$ with
$\hat{h}_\Lambda(k)=1, k^2 \leq \Lambda^2 \mbox{ and } \hat{h}_\Lambda(k)=0,
k^2 \geq 2\Lambda^2.$
The point of all this is that one might just as well assume that $A_\mu$ in (1.1)
is $C^\infty$ and polynomial bounded to begin with.  If one succeeds in 
calculating a useful determinant one can then replace the potential in 
$F_{\mu\nu}$ with $A_\mu^\Lambda$ before the final functional 
integration over the
gauge field.  Or one may select some other regularization procedure.

Essentially we are instructed to integrate over all potentials, which requires
knowledge of the determinant for all fields.  What all fields means depends on
the dimensionality of spacetime.  In Euclidean space we need the determinant for
fields $\mathbf{B}$ and $\mathbf{E}$ in four dimensions, $\mathbf{B}$ in three
dimensions, and a unidirectional magnetic field $B$ in two dimensions.  We have
shown in [1] that an integral of the fermionic determinant in
$\mbox{QED}_2$ over the
fermion's mass gives the determinant in $\mbox{QED}_4$  for the field 
$\mathbf{B}=(0,0,B(x,y))$.  It will be shown in Section 2 that the determinant
in $\mbox{QED}_3$ may be calculated in the same way for this $\mathbf{B}\mbox{-field}.$
And we will show in Section 3
that a mass integral of the fermionic determinant in $\mbox{QED}_3$ gives the determinant in $\mbox{QED}_4$ for a static, directionally-varying
magnetic field $\mathbf{B}(\mathbf{r})$.

The author has repeatedly encountered the assertion that the fermionic determinant of $\mbox{QED}_2$ is known explicitly.  This is true for the case of massless
fermions---the Schwinger model [9]---but not for the all-important case of 
massive fermions considered here.  We note in passing that there is
 evidence that
the massive fermionic determinant in $\mbox{QED}_2$ is discontinuous
 at $m=0$ for 
magnetic fields with nonvanishing flux [3]. This would imply that the
Schwinger model's fermionic determinant cannot in general be obtained as
the zero-mass limit of $\mbox{QED}_2\mbox{'s}$.

As the representation (1.1) makes clear,  the calculation of a fermionic 
determinant is really just a problem in quantum mechanics involving the 
calculation of the energy levels and their degeneracy of the Pauli operator
\begin{eqnarray}
( \Psl-\hspace{-0.5em}\Asla)^\dagger( \Psl - \hspace{-0.5em}\Asla)=
(P-A)^2 +\frac{1}{2}
\sigma^{\mu\nu}F_{\mu\nu} \geq 0.
\end{eqnarray}
Since the determinant is required for general fields, probably the best that can
be done at present is to make estimates that place stringent bounds on the
determinant.  Inevitably it is the Zeeman term $\sigma F$ that complicates
matters.  If  it is simply ignored then the zero modes of the Pauli operator
are absent, thereby causing an unacceptable modification of the infrared 
behavior of QED.

It is by now a piece of folklore that the Pauli operator in two space dimensions
in a unidirectional magnetic field $ B \rightarrow 0$ at infinity
has associated
eigenvalues with finite degeneracy.  This is necessary if one is to make sense
out of the trace in (1.1) or any other definition of a determinant the author is
aware of.  This question has been discussed in [1,3,4].  We know in particular 
that polynomial, infinite flux, unidirectional magnetic fields are associated 
with infinite degeneracy [10].  Whether infinite flux in general implies an 
infinitely degenerate ground state is not known.  Some results in this direction
are given in [11].  Here we will consider only those unidirectional magnetic
fields with finite flux, which is consistent with the introduction of a 
volume cutoff and which is required to define QED before taking
the thermodynamic limit.

Before listing the known bounds on the determinants, including those obtained
here, we mention two analytic calculations for finite flux fields: the
determinant in $\mbox{QED}_2$ for the radially symmetric cylindrical field [3],
\begin{eqnarray}
B(r) = \frac{\Phi}{2\pi} \frac{\delta (r - a) }{a},
\end{eqnarray}
and the determinant in $\mbox{QED}_3$ for the field [12]
\begin{eqnarray}
B(x,y) = \frac{B}{[\cosh(x/\lambda)]^2},
\end{eqnarray}
localized in a strip of finite extent in the y-direction.

Table I summararizes the known bounds on the fermionic determinants in QED.
The lower bounds are for the fields $\mathbf{B}=(0,0,B(x,y)) \mbox{, } B(x,y) \geq 0 \mbox{ or }
B(x,y) \leq 0$.  The lower bound for $\mbox{QED}_3$ is new and will be dealt
with in Section 2.
The upper bound on $\mbox{QED}_4\mbox{'s}$ determinant for a static, square integrable,
directionally varying magnetic field $\mathbf{B}(\mathbf{r})$, where \bfrs 
is a point in $\R^3$, is also new and will be established in Section 3.  The other
bounds have been previously derived.  While the bounds for $\mbox{QED}_{2,3}$ indicate
stability, the lower bound for $\mbox{QED}_4$, for the class of static magnetic fields
considered, indicates that the contribution of the virtual fermion currents to the
effective energy at the one-loop level is unbounded from below as the field's flux
is increased.  As noted above,  it is the one-loop effective action, or energy in
the special case of static fields in Euclidean space,  that is revelant to the
nonperturbative analysis of QED.  Section 3.3 is devoted to establishing bounds
on the induced spin in planar QED with finite mass fermions in the presence of
inhomogeneous background magnetic fields.

Finally, we would like to comment on the case of general static fields
$\mathbf{B}(\mathbf{r})$ and $\mathbf{E}(\mathbf{r})$ in $\mbox{QED}_4$.
It seems to be taken for granted that the effective Lagrangian for
constant $\mathbf{B}$ and $\mathbf{E}$ [5,14] is an indication of the
behavior of the fermionic determinant for general fields, provided one
accepts the fudging of the thermodynamic limit involved.  Now it is well
known that $F_{\mu\nu}$ can be reduced to block diagonal form for
constant fields by two rotations in $\R^4$ (corresponding to a Lorentz
boost and a rotation in Minkowski space).  As a result the constant
field case reduces to the calculation of the spectrum of two uncoupled
harmonic oscillators describing the planar motion of two independant
charged particles in the normal magnetic and electric (in the Euclidean
sense) fields $\frac{1}{2}(| \mathbf{B} + \mathbf{E}| \pm | \mathbf{B} -
\mathbf{E} |)$.  Therefore, constant fields are not generic in any sense,
and the completly open problem of general static fields may yet prove to
be of substantial interest.

% +++

\section{THREE-DIMENSIONAL QED}

\subsection{Connection between the fermionic determinants in QED$_3$ and QED$_2$}

We choose for the Dirac matrices in three dimensions the $2\times 2$ matrices
$\gamma^\mu= (i \sigma_1, i \sigma_2, i \sigma_3),$ where the $\sigma_i\mbox{'s}$
are the Pauli matrices.  In this case definition (1.1) of the determinant in $\mbox{QED}_3$
reduces to
\begin{eqnarray}
\ln\mbox{det}_{QED3}=\frac{1}{2} \int^\infty_0 \frac{dt}{t} \mbox{Tr} \left ( e^{-P^2t}- \exp \{
-[(\mathbf{P}-\mathbf{A})^2 - \boldsymbol{\sigma}\cdot\mathbf{B}]t\}\right )e^{-tm^2}. 
\end{eqnarray}

This definition (and regularization) of the fermionic determinent is parity 
conserving and gives no Chern-Simons term, which is known to be regularization
dependent [15].  Such a term may always be added.  In order to relate
$\mbox{det}_{QED3}$
to Euclidean QED in two dimensions let $\bfB=(0,0,B(\bfr)),$ 
$\mathbf{A}=(\mathbf{A_\perp(r)},0)$ and $\mathbf{A}_\perp=(A_x,A_y)$, where \bfrs is a 
point in the $xy\mbox{-plane}$.  Enclosing the $z\mbox{-axis}$ (which may also
be called the time axis) in a large box of length $Z$ we get
\begin{eqnarray}
\ln\mbox{det}_{QED3}(m^2) & = & \frac{Z}{4\pi^{1/2}} \int_0^\infty
\frac{dt}{t^{3/2}} \mbox{Tr} \left( e^{-\mathbf{P}_\perp^2t}  \right. \nonumber \\
 &   & \left.\vphantom{e^{-\mathbf{P}_\perp^2t}} -exp \left\{ - [( \mathbf{P}_\perp - \mathbf{A}_\perp)^2 -
 \sigma_3B]t\right\} \right) e^{-tm^2}, 
\end{eqnarray}
where we used
\begin{eqnarray}
\mbox{Tr}_{space}\; e^{-P^2_3t}= \frac{Z}{(4\pi t)^{1/2}};
\end{eqnarray}
the remaining trace in (2.2) is over space and spin indices.

The fermionic determinant in Euclidean $\mbox{QED}_2$ (denoted by
$\mbox{det}_{Sch}$ in refs.\
[1--4]) is
\begin{eqnarray}
\ln\mbox{det}_{QED2}(m^2) &=& \frac{1}{2} \int^\infty_0 \frac{dt}{t} \mbox{Tr} 
\left( e^{-\mathbf{P}_\perp^2 t} \right. \nonumber  \\
 & & \left.\vphantom{e^{-\mathbf{P}-\perp^2 t}} 
-exp \left\{ -[(\mathbf{P}_\perp - \mathbf{A}_\perp)^2 -
 \sigma_3B] \right\} \right) e^{-tm^2}.
\end{eqnarray}
Using nothing more than $\int^\infty_0 dE \exp(-tE^2)=(\pi/4t)^{1/2}$
we get the connection between the two determinants:
\begin{eqnarray}
\ln\mbox{det}_{QED3}(m^2) &=& \frac{Z}{\pi} \int^\infty_0 dE
\ln\mbox{det}_{QED2}(E^2 + m^2) \nonumber \\
& =  & \frac{Z}{2\pi} \int^\infty_{m^2} \frac{dM^2}{\sqrt{M^2-m^2}}
\ln\mbox{det}_{QED2}(M^2).
\end{eqnarray}
As for $B$, we are assuming that it is a smooth, polynomial bounded
$C^\infty$ function
with finite flux; it will also be assumed to be square integrable.

As a check on (2.5) one may substitute the second-order contribution to
$\mbox{QED}_2\mbox{'s}$ determinant obtained by expanding (2.4),
\begin{eqnarray}
\ln\mbox{det}_{QED2} =- \frac{1}{2\pi} \int \frac{d^2k_\perp}{(2\pi)^2}
|\hat{B}(\mathbf{k}_\perp)|^2 \int^1_0 dz
\frac{z(1-z)}{z(1-z)k_\perp^2+m^2}+O (B^4),
\end{eqnarray}
and obtain the canonical result
\begin{eqnarray}
\ln\mbox{det}_{QED3} & =- \frac{Z}{4\pi} \int \frac{d^2k_\perp}{(2\pi)^2}
|\hat{B}(\mathbf{k}_\perp)|^2 \int^1_0 dz
\frac{z(1-z)}{[z(1-z)k_\perp^2+m^2]^{1/2}}+O (B^4),
\nonumber \\
\end{eqnarray}
for the unidirectional field B$(\bfr)$.

An immediate consequence of (2.5) is that the ``diamagnetic'' bound in QED$_2$
[8,13],
\begin{eqnarray}
\mbox{det}_{QED2} \leq 1,
\end{eqnarray}
implies a ``diamagnetic'' bound in $\mbox{QED}_3$ for the static field $\bfB=(0,0,B(\bfr))$,
\begin{eqnarray}
\mbox{det}_{QED3} \leq 1.
\end{eqnarray}
The term ``diamagnetic'' is placed in quotation marks as it is really an expression
of the paramagnetic property of fermions as definitions (2.1) and (2.4) make clear.

\subsection{Lower bound on $\boldsymbol{\ln}$ det$_{\mathbf{QED3}}$}

It is now possible to obtain a lower bound on $\ln\mbox{det}_{QED3}$
with the aid of
(2.5) for the field $\bfB=(0,0,B(\bfr))$, where $B(\bfr) \geq 0$ or $B(\bfr) \leq 0$,
and $\bfr$ is a point in the $xy\mbox{-plane}$.
For $B(\bfr) \geq 0$ we showed in [4] that
\begin{eqnarray}
\ln\mbox{det}_{QED2} \geq \frac{1}{4\pi} \int d^2r \left[ B(\bfr) -
(B(\bfr)+m^2) \ln \left( 1 + \frac{B(\bfr)}{m^2} \right) \right] .
\end{eqnarray}
For $B \leq 0$, simply replace $B$ with $-B$.  Thus, (2.5) and (2.10) give
\begin{eqnarray}
\ln\mbox{det}_{QED3} & \geq & \frac{Z}{8\pi^2} \int^{\infty}_{m^2} 
\frac{dM^2}{\sqrt{M^2-m^2}} \int d^2r
\left[\vphantom{\frac{B(\mathbf{r})}{M^2}} B(\bfr) \nonumber \right.  \nonumber \\
&   & \left. -(B(\bfr) +M^2) \ln \left ( 1 + \frac{B(\bfr)}{M^2} \right )
\right ] \nonumber \\
& = & \frac{Z|m|^3}{6\pi} \int d^2r \left [1+ \frac{3B(\bfr)}{2m^2} -
\left ( 1+ \frac{B(\bfr)}{m^2} \right )^{3/2} \right ].
\end{eqnarray}
This is our main bound.  A simpler, less stringent bound can be obtained by 
noting that
\begin{eqnarray}
1 + \frac{3}{2}x - (1+x)^{3/2} \geq -x^{3/2}, \:\:\:    x \geq 0,
\end{eqnarray}
in which case
\begin{eqnarray}
\ln\mbox{det}_{QED3} \geq - \frac{Z}{6\pi} \int d^2r \; |B(\bfr)|^{3/2}.
\end{eqnarray}
The absolute value has been added to include the two possible signs of $B$. 
Note that this bound is uniform in the fermion's mass.

As a formal check on (2.13) we can compare it with Redlich's [16] result for the
zero-mass limit of $\ln\mbox{det}_{QED3}$ for the case of a constant magnetic field.  Combining his Eq.\ (4.25) with (2.13) requires
\begin{eqnarray}
\lim_{m\rightarrow 0}\ln\mbox{det}_{QED3} = - \frac{V\zeta (3/2)}{4\pi^2\sqrt{2}}B^{3/2} 
\geq - \frac{V}{6\pi}B^{3/2},
\end{eqnarray}
or $\zeta(3/2) \leq 2^{3/2} \pi/3$, where $\zeta (3/2)$ is the Riemann zeta 
function
\begin{eqnarray}
\zeta (3/2) = \sum^\infty_{n=1} n^{-3/2}, 
\end{eqnarray}
and $V$ is the volume of a large box in $\R^3$.  Since $\zeta(3/2)=2.612$ to 
four significant figures, (2.14) implies $2.612 \leq 2.962$.  

If the $z\mbox{-axis}$ is relabeled as the time axis then the effective one-loop
energy $\mathcal{E}$ of $QED_3$ is bounded by
\begin{equation}
0 \leq \mathcal{E} \leq \frac{1}{6\pi} \int d^2r \; |B(\mathbf{r})|^{3/2},
\end{equation}
where the lower bound comes from the diamagnetic bound (2.9).  Hence,
our results support stability for the class of static magnetic fields
considered here.

As another check on our results consider the field of ref.\ [12] given
by (1.5).  Eq.\ (2.16) gives the bound
\begin{equation}
0 \leq \mathcal{E} \leq L\lambda(eB)^{3/2}/12,
\end{equation}
where the coupling $e$ has been restored, and $L$ is the length of the
strip in the $y\mbox{-direction}$.  The authors of ref.\ [12] calculated
$\mathcal{E}$ analytically.  The zero-mass limit of $\mathcal{E}$, given
by their Eq.\ (22), allows a direct check on (2.17):
\begin{equation}
\mathcal{E} = \frac{L \lambda (eB)^{3/2}}{8 \sqrt{2} \pi} \left [\zeta(3/2)
-\frac{15}{16\pi}\zeta(5/2)\frac{1}{eB\lambda^2}+\cdots\right ].
\end{equation}
Thus, (2.17) and (2.18) give $0.073-\cdots \leq 0.083$.

\subsection{Induced spin}

Using the above results we can obtain a lower bound on the spin induced
in the vacuum by a static, unidirectional magnetic field for all finite
values of the fermion mass.  In $2+1$ dimensions the normal ordered spin
density in the field $\mathbf{B}(\mathbf{r})=(0,0,B(\mathbf{r}))$
derived from the potential $\mathbf{A} = (\mathbf{A}_\perp
(\mathbf{r}),0)$ is given by 
\begin{eqnarray}
S(\mathbf{r}; B) & = & \frac{1}{2} \bigl<[ \psi^\dagger (\mathbf{r},t)
\frac{1}{2} \sigma_3, \psi(\mathbf{r},t) ]_-\: \bigr> \nonumber \\
& = & -  \frac{1}{4} \lim_{\epsilon\downarrow 0} \sum_{n} \int_C
\frac{d\omega}{2\pi i} e^{-i\omega\epsilon} \psi^\dagger_n (\mathbf{r})
\,\sigma_3\,\psi_n (\mathbf{r}) \nonumber \\
&   & \times \left[ (E_n-\omega)^{-1} + (E_n+\omega)^{-1} \right],
\end{eqnarray}
where the contour $C$ runs below the negative real $\omega\mbox{-axis}$,
passes through the origin and continues running above the positive real
$\omega\mbox{-axis}$. The $\psi_n$ are energy eigenstates
\begin{eqnarray}
H\psi_n & = & E_n \psi_n  \nonumber \\
H & = & \boldsymbol{\alpha}\cdot(\mathbf{p}-\mathbf{A}_\perp) + \beta m,
\end{eqnarray}
with $\gamma^1=i\sigma_1, \ \gamma^2=i\sigma_2, \ \beta=\sigma_3, \ \mbox{and }
\boldsymbol{\alpha}=\beta\boldsymbol{\gamma}$.
Then
\begin{eqnarray}
S(\mathbf{r};B) & = & - \frac{1}{4} \lim_{\epsilon\downarrow 0} \int_C
\frac{d\omega}{2\pi i} e^{-i\omega\epsilon} \; \mbox{tr}\,
\left< \mathbf{r} | ( \Psl_\perp -
\hspace{-0.5em}\Asla_\perp +m -\omega\sigma_3)^{-1} \right. \nonumber \\
&   & \left. + (\Psl_\perp -\hspace{-0.5em}\Asla_\perp +m +\omega\sigma_3)^{-1} \,|
\,\mathbf{r}
\right> \nonumber \\
& = & -\frac{m}{2} \lim_{\epsilon\downarrow 0} \int_C
\frac{d\omega}{2\pi i} e^{-i\omega\epsilon} \nonumber \\
&   & \times\:\mbox{tr} \left< \mathbf{r}| \left( 
(\mathbf{P}_\perp - \mathbf{A}_\perp)^2 - \sigma_3 B +m^2
-\omega^2 \right)^{-1} | \mathbf{r} \right>.
\end{eqnarray}
Now rotate the $\omega\mbox{-contour } 90^o$ counterclockwise while letting
$\epsilon \rightarrow -i\epsilon$ to effect a Wick rotation.  This gives
\begin{equation}
S(\mathbf{r},B) = -m \int^{\infty}_0 \frac{dE}{2\pi} \mbox{tr} \left<
\mathbf{r} | ((\mathbf{P}_\perp - \mathbf{A}_\perp)^2-\sigma_3 B
+m^2 +E^2)^{-1} | \mathbf{r} \right>.
\end{equation}
To make sense out of this the spin density at B$=0$ has to be subtracted
out.  Changing the integration variable to $M^2=E^2+m^2$ and integrating
over the $xy\mbox{-plane}$, we obtain the total induced spin,
\begin{eqnarray}
S(B)-S(0) & = & -\frac{m}{4\pi} \int^\infty_{m^2} \frac{dM^2}{\sqrt{M^2-m^2}} \,
\mbox{Tr} \left[((\mathbf{P}_\perp - \mathbf{A}_\perp)^2
-\sigma_3B+M^2)^{-1}  \right.  \nonumber \\
&  & \left. \ \ \ \ \ \ \ \ \ \ \ \ \ \ \ \ \ \ \ \ \ \ \ \ \ \ \ \ \ \ \
-(P_\perp^2+M^2)^{-1}\right].
\end{eqnarray}

We will now relate the induced spin to the determinants 
$\mbox{det}_{QED3}$ and $\mbox{det}_{QED2}$.
From (2.4),
\begin{eqnarray}
\frac{\partial}{\partial m^2}\ln\mbox{det}_{QED2}& =\frac{1}{2}
\mbox{Tr} \left[((\mathbf{P}_\perp - \mathbf{A}_\perp)^2
-\sigma_3B+M^2)^{-1}-(P^2_\perp+M^2)^{-1}\right],
\nonumber \\
\end{eqnarray}
and hence,
\begin{eqnarray}
S(B)-S(0)=-\frac{m}{2\pi} \int^\infty_{m^2} \frac{dM^2}{\sqrt{M^2-m^2}}
\frac{\partial}{\partial M^2} \ln\mbox{det}_{QED2}\: (M^2).
\end{eqnarray}
Since $\mbox{det}_{QED2}$ is even in $B$ so is the induced spin.

From (2.2), after relabeling the $z\mbox{-axis}$ the time axis, we get
\begin{eqnarray}
\frac{\partial}{\partial m} \ln\mbox{det}_{QED3} & = & 
\frac{mT}{2\pi^{1/2}}
\int^{\infty}_0 \frac{dt}{t^{1/2}} \mbox{Tr} 
\left(\vphantom{-e^{-P_\perp^2t}}\exp \left\{
-[(\mathbf{P}_\perp - \mathbf{A}_\perp)^2
- \sigma_3B]t\right\}
\right.  \nonumber \\
&   & \left. \ \ \ \ \ \ \ \ \ \  \ \ \ \ \ \ \ \ \ \ \ \ \ \ \ \ \ \ \ \ 
-e^{-P_\perp^2t} \right) e^{-tm^2}.
\end{eqnarray}
Again using $t^{-1/2}=(4/\pi )^{1/2} \int^\infty_0 dE \exp (-tE^2)$ and
then changing the integration variable to $M^2=E^2+m^2$ gives
\begin{eqnarray}
\frac{\partial}{\partial m} \ln\mbox{det}_{QED3} & = & \frac{mT}{2\pi}
\int^\infty_{m^2} \frac{dM^2}{\sqrt{M^2-m^2}} \mbox{Tr} \Bigl[ (
(\mathbf{P}_\perp-\mathbf{A}_\perp)^2 - \sigma_3B+M^2)^{-1} \Bigl. \nonumber \\
&   & \Bigr. \ \ \ \ \ \ \ \ \ \ \ \ \ \ \ \ \ \ \ \ \ \ \ \ \ \ \  
-(P_\perp^2+M^2)^{-1} \Bigr].
\end{eqnarray}
Comparing (2.27) with (2.23) gives
\begin{eqnarray}
\frac{\partial}{\partial m} \ln\mbox{det}_{QED3} = -2T \; [ S(B)-S(0)], 
\end{eqnarray}
which is what one expects from formal manipulation of the fermionic
functional integral for $\mbox{det}_{QED3}$.

We are now in a position to obtain bounds on the induced spin.  From
Eqs.\ (5)--(6) in [4],
\begin{eqnarray}
\frac{\partial}{\partial m^2} \ln\mbox{det}_{QED2} \leq \frac{\Phi}{4\pi m^2}
-\frac{1}{4\pi} \int d^2r \ln \left(1 +
\frac{B(\mathbf{r})}{m^2}\right),
\end{eqnarray}
where it is again assumed that $B(\mathbf{r}) \geq 0$ or $B(\mathbf{r})
\leq 0$ with $\mathbf{r}$ a point in the $xy\mbox{-plane}$ and $\Phi =
\int d^2r B(\mathbf{r})$.  Substituting (2.29) in (2.25) gives, for $m >
0$,
\begin{eqnarray}
S(B)-S(0) & \geq & \frac{m}{8\pi^2} \int^{\infty}_{m^2}
\frac{dM^2}{\sqrt{M^2-m^2}} \Bigl[\int d^2r \ln \Bigl(
1+\frac{B(\mathbf{r})}{M^2}\Bigr) - \frac{\Phi}{M^2} \Bigr] \nonumber \\
& = & \frac{m}{4\pi} \int d^2r \Bigl[(B(\mathbf{r})+m^2)^{1/2} -
\frac{B(\mathbf{r})}{2m}-m\Bigr],
\end{eqnarray}
while for $m < 0$
\begin{eqnarray}
S(B)-S(0) \leq -\frac{|m|}{4\pi} \int d^2r \Bigl[(B(\mathbf{r})+m^2)^{1/2}  -
\frac{B(\mathbf{r})}{2|m|}-|m|\Bigr].
\end{eqnarray}
For $B \leq 0$ simply change the sign of $B$ in (2.30)--(2.31).
Elementary estimates indicate that the integrals in (2.30)--(2.31)
converge if $B \in L^2(\R^2)$.

Of particular interest is the $m=0$ limit of the induced spin since this
is related to the vacuum condensate $<\overline{\psi}\psi>_{m=0}$ in the
presence of an inhomogeneous background magnetic field.  If the range of
$B$ is finite and independant of $m$, then the $m=0$ limit may be safely
taken, giving
\begin{eqnarray}
\bigl[S(B)-S(0)\bigr]_{m\downarrow 0} \geq - \Phi/8\pi,
\end{eqnarray}
and
\begin{eqnarray}
\bigl[S(B)-S(0)\bigr]_{m\uparrow 0} \leq  \Phi/8\pi.
\end{eqnarray}
These limits are consistent with the results of Parwani [17].
Comparing (2.28) with (2.32)--(2.33) it is evidently possible for
$\ln\mbox{det}_{QED3}$ to have a discontinuous mass derivative at $m=0$.

Finally, the vacuum condensate for the magnetic field
\begin{eqnarray}
B(r) = B(1+r^2/R^2)^{-2},
\end{eqnarray}
has been calculated by Dunne and Hall [18].  Assume B$>0$.  Since the
authors use $4\times 4 \mbox{ } \gamma\mbox{-matrices}$, their result has
to be divided by 4 to correct for this and to relate their condensate to
the spin density. Their Eq.\ (29) then gives
\begin{eqnarray}
\bigl[S(\mathbf{r};B)-S(\mathbf{r};0)\bigr]_{m\rightarrow 0} =
-\mbox{sign}(m)\Bigl(1-\frac{2\pi}{\Phi}\Bigr)\frac{B(r)}{8\pi},
\end{eqnarray}
and hence
\begin{eqnarray}
\bigl[S(B)-S(0)\bigr]_{m\rightarrow 0} =
-\mbox{sign}(m)\Bigl(1-\frac{2\pi}{\Phi}\Bigr)\frac{\Phi}{8\pi},
\end{eqnarray}
The $m=0$ limit of the right-hand sides of (2.30)--(2.31) give the same
results as in (2.32)--(2.33).  Our results and those of ref.\ [18] are
therefore consistent.

% +++

\section{FOUR-DIMENSIONAL QED}

\subsection{Connection between the fermionic determinants in QED$_3$ and
QED$_4$ }

In order to make the connection we choose the static potential
$A_{\mu}=(0,\mathbf{A}(\mathbf{r}))$.  It is assumed that $\mathbf{A}$
is polynomial bounded, $C^\infty$ and that $\mathbf{A} \in L^3(\R^3)$.
Why $\mathbf{A}$ is chosen to be in $L^3$ will be explained below (see
also end of Section 3.3.  We will require that the magnetic field
$\mathbf{B}=\mathbf{\nabla}\times\mathbf{A}$ be square
integrable.  If $\mathbf{B} \in L^2(\R^3)$ and $\mathbf{A}$ is also
assumed to be in the Coulomb gauge
$\mathbf{\nabla}\cdot\mathbf{A}=0$, then by
Sobolev-Talenti-Aubin inequality [19],
\begin{eqnarray}
\int d^3r \mathbf{B}(\mathbf{r})\cdot\mathbf{B}(\mathbf{r}) \geq
(27\pi^4/16)^{1/3}
\sum^3_{i=1}\left( \int d^3r |A_i(\mathbf{r})|^6\right)^{1/3},
\end{eqnarray}
that is, $\mathbf{A}\in L^6(\R^3)$ as well as $L^3(\R^3)$.  As a simple
consequence of this and H\"{o}lder's inequality [20], 
\begin{eqnarray}
\|fg\|_r & \leq & \|f\|_p\|g\|_q \nonumber \\
p^{-1}+q^{-1} & = & r^{-1}, \mbox{ } 1 \leq p,q,\: r \leq \infty,
\end{eqnarray}
$\mathbf{A} \in\underset{3\leq p \leq 6}\cap L^p (\R^3)$.  No assumption has to
be made about finite flux as it is always zero.  Finally, we choose the
chiral representation of the $\gamma\mbox{-matrices}$ so that 
$\sigma_{ij}=\bigl( \begin{smallmatrix} 
-\sigma_k&0\\ 0&-\sigma_k
\end{smallmatrix} \bigr)
$ with $i,j,k=1,2,3$ in cyclic order.

Following these preliminaries, Eq.\ (1.1) for $\mbox{QED}_4\mbox{'s}$ fermionic
determinant reduces to 
\begin{eqnarray}
\ln\mbox{det}_{ren} & = & \frac{T}{2} \int^\infty_0 \frac{dt}{t} \left[
\frac{2}{(4\pi t)^{1/2}} \mbox{Tr} \left( e^{-\mathbf{P}^2t} \right. \right. 
\nonumber \\
&   & \left. \left.\vphantom{e^{-P^2t}}  -\exp \left\{
-[(\mathbf{P} - \mathbf{A})^2 - \boldsymbol{\sigma}\cdot\mathbf{B}]t \right\}
\right) + \frac{\|\mathbf{B}\|^2}{12\pi^2} \right] e^{-tm^2},
\end{eqnarray}
where $T$ is the dimension of the time box and $\|\mathbf{B}\|^2= \int
d^3 r \mathbf{B}(\mathbf{r})\cdot\mathbf{B}(\mathbf{r})$.  We have 
used (2.3) again for $\mbox{Tr}_{space}\,e^{-P^2_0 t}$, 
exchanging $Z$ for $T$.
The factor 2 in (3.3) comes from a partial spin sum; the remaining spin
trace is over a two-dimensional space.  The determinant in $\mbox{QED}_3$ in
the presence of $\mathbf{B}(\mathbf{r})$ is, from (1.1), 
\begin{eqnarray}
\ln \mbox{det}_{QED3}(m^2) & = & \frac{1}{2} \int^\infty_0 \frac{dt}{t} \mbox{Tr}
\left( e^{-\mathbf{P}^2t} \right. \nonumber \\
&   &  \left.\vphantom{e^{-P^2t}} -exp \left\{ -[(\mathbf{P}-\mathbf{A})^2 -
\boldsymbol{\sigma}\cdot\mathbf{B}]t \right\} \right) e^{-tm^2}. 
\end{eqnarray}
% is the sigma in boldface.
Thus, we get the connection between QED$_3$ and QED$_4$ for static magnetic
fields:
\begin{eqnarray}
\ln\mbox{det}_{ren} & = & \frac{2T}{\pi} \int^\infty_0 dE \left(
\ln\mbox{det}_{QED3} (E^2+m^2) + \frac{\|\mathbf{B}\|^2}{24\pi^{3/2}}
\int^\infty_0 \frac{dt}{t^{1/2}} e^{-(E^2+m^2)t} \right) \nonumber \\
 & = & \frac{T}{\pi} \int^{\infty}_{m^2} \frac{dM^2}{\sqrt{M^2-m^2}}
\left( \ln\mbox{det}_{QED3}(M^2)+\frac{\|\mathbf{B}\|^2}{24\pi\sqrt{M^2}}
\right).
\end{eqnarray}

In order to get the upper bound on $\ln\det_{ren}$ in Table I it is
useful to isolate the second-order contribution to $\ln\det_{QED3}$.
Denoting the remainder by $\ln\det_4$ we get
\begin{eqnarray}
\ln\mbox{det}_{QED3}(1-S\hspace{-0.5em}\Asla) & = & - \frac{1}{4\pi}
\int \frac{d^3k}{(2\pi)^3}\:
|\hat{\mathbf{B}}(\mathbf{k})|^2 \int^1_0 dz
\frac{z(1-z)}{[z(1-z)k^2+m^2]^{1/2}} \nonumber \\
&  &  \ \ \ \ \ \ \ \ \ \ \ \ \ \ \ \ \ \ \ \ \ \ \ \ \ \ \
+ \ln\mbox{det}_{4} (1-S\hspace{-0.5em}\Asla).
\end{eqnarray} 
The first term on the right-hand side of (3.6) was obtained by expanding
(3.4) to second order.  Graphically, $\ln\mbox{det}_4$ is the sum of all even
order one-loop fermion graphs in three dimensions, beginning with the
fourth-order box graph since definition (3.4) respects Furry's 
theorem or $C\mbox{-invariance}$.  Thus, restoring e,
\begin{equation}
\ln\mbox{det}_4(1-eS\hspace{-0.5em}\Asla)=-\sum^\infty_{n=4} \frac{e^n}{n}
\mbox{Tr}(S\hspace{-0.5em}\Asla)^n.
\end{equation}
The operator $S\hspace{-0.5em}\Asla$ is a bounded operator on $L^2 (\R^3,
\sqrt{k^2+m^2} d^3k; \C^2)$ for $\mathbf{A} \in L^p(\R^3)$ for
some $p>3$, which is the case here. In addition,
$S\hspace{-0.5em}\Asla=(\Psl+m)^{-1}\hspace{-0.5em}\Asla(X)$ belongs
to the trace ideal
$\mathcal{C}_p$ for $p > 3 \  (\mathcal{C}_p= \{A| \; \|A\|^n_n \equiv \mbox{Tr}
((A^\dagger A)^{n/2}) < \infty \} )$ [6, 21-23]. 
As a result, the eigenvalues $1/e_n$ of
the compact operator $S\hspace{-0.5em}\Asla$
(none of which are real for $m \neq 0$;
see Section 3.3 are such that $\sum_{n=1}^\infty |e_n|^{-p} < \infty $
[24].  Therefore, the series in (3.7) has a finite radius of
convergence, although our analysis will not rely on this.  More will be
said about $\det_4$ for general $e$ in Section 3.3. For the present, note 
that it is defined for all real $e$ by (3.4) and (3.6) (see Eq.\ (3.14) 
below).  But already one begins to see the 
usefulness of $\mathbf{A} \in L^3(\R^3)$.

Inserting (3.6) in (3.5) gives
\begin{eqnarray}
\ln\mbox{det}_{ren} & = & \frac{T}{4\pi^2} \int \frac{d^3k}{(2\pi)^3}
|\hat{\mathbf{B}}(\mathbf{k})|^2 \; \int^1_0 dz z(1-z) \ln 
\left[ \frac{z(1-z)k^2+m^2}{m^2} \right] \nonumber \\
 &   & + \frac{T}{\pi} \int^{\infty}_{m^2} \frac{dM^2}{\sqrt{M^2-m^2}}
\ln\mbox{det}_4(M^2).
\end{eqnarray}
The first term on the right-hand side is the second-order vacuum
polarization contribution to $det_{ren}$ in a static magnetic field,
renormalized at zero momentum transfer.

\subsection{Upper bound on $\boldsymbol{\ln}$ det$_{ren}$}

An upper bound can be put on $\ln\det_{ren}$ in a general static
magnetic field $\mathbf{B}(\mathbf{r})$ with the help of (3.8)
and the diamagnetic inequality for QED$_3$ [8],
\begin{eqnarray}
| \mbox{det}_{QED3} (1 -eS\hspace{-0.5em}\Asla)| \leq 1,
\end{eqnarray}
where $\mathbf{A}$ is the smooth potential we introduced in Section 3.1.
For $m \neq 0 \ \mbox{det}_{QED3}$ has no zeros for real $e$ (see Section 3.3)
and if det$_{QED3}|_{e=0}=1$, then we can write instead of (3.9),
\begin{eqnarray}
0 < \mbox{det}_{QED3} (1-eS\hspace{-0.5em}\Asla) \leq 1. 
\end{eqnarray}
A few comments on (3.9) and (3.10) are in order.  The diamagnetic
inequality is general and follows for any determinant that is 
obtained as the continuum limit of a lattice theory obeying 
reflection positivity.  On the lattice Wilson fermions may be used,
and since they are $CP$ invariant, there is no Chern-Simons term [25].
The fact that the continuum limit of det$_{QED3}$ coincides with
definition (3.4) follows from Seiler's Statement 5.4 and his Eq.\
(7.20) (without the counterterm which is not needed in QED$_3$) in 
ref.\ [6].

Now (3.10) and (3.6) imply that
\begin{eqnarray}
\ln\mbox{det}_4 (1-S\hspace{-0.5em}\Asla)
\leq \frac{1}{4\pi} \int \frac{d^3k}{(2\pi)^3}
|\hat{\mathbf{B}}(\mathbf{k})|^2 \int^1_0 dz
\frac{z(1-z)}{[z(1-z)\,k^2+m^2]^{1/2}}.
\end{eqnarray}
This remarkable consequence of the paramagnetism of charged 
spin-$\frac{1}{2}$ fermions implies
that all the nonlinearities of $\ln\det_4$ are bounded by a quadratic in
the magnetic field.  Inserting (3.11) into (3.8) gives, for 
$\|\mathbf{B}\|^2 \geq |m|$  (restoring $e$, recall that
$e^2\|\mathbf{B}\|^2=e^2 \int d^3r \mathbf{B}\cdot\mathbf{B}$ has the
dimension of mass in both three and four dimensions):
\begin{eqnarray}
\ln\mbox{det}_{ren} & \leq & \frac{T}{4\pi^2} \int \frac{d^3k}{(2\pi)^3}
|\hat{\mathbf{B}}(\mathbf{k})|^2 \int^1_0 dz \: z(1-z) \ln
\left[\frac{z(1-z)k^2+m^2}{m^2}\right]  \nonumber \\
&   & + \frac{T}{4\pi^2} \int^{ \| \mathbf{B} \| ^4}_{m^2}
\frac{dM^2}{\sqrt{M^2-m^2}} \int \frac{d^3k}{(2\pi)^3} | \hat{\mathbf{B}}
(\mathbf{k})|^2 \int^1_0 dz \frac{z(1-z)}{[z(1-z)k^2+M^2]^{1/2}}  \nonumber \\
&   & + \frac{T}{\pi} \int_{ \| \mathbf{B} \|^4}^{\infty}
\frac{dM^2}{\sqrt{M^2-m^2}} \ln\mbox{det}_4(M^2) \nonumber \\  
& \leq & \frac{T}{4\pi ^2} \int \frac{d^3k}{(2\pi)^3}
| \hat{\mathbf{B}}(\mathbf{k}) |^2 \int^1_0 dz \: z(1-z) \ln \left[
\frac{4 \| \mathbf{B} \| ^4+2z(1-z)k^2-2m^2}{m^2} \right] \nonumber \\
&   & + \frac{T}{\pi} \int^\infty_{ \| \mathbf{B} \| ^4 } 
\frac{dM^2}{\sqrt{M^2-m^2}} \ln\mbox{det}_4(M^2).
\end{eqnarray}
The argument of the logarithm the last line of (3.12) has been
simplified somewhat using $2\sqrt{xy} \leq x+y \mbox{ for }
x,y \geq 0$ .

The last term in (3.12) can be estimated for strong fields. Thus let
$\mathbf{B} \rightarrow \lambda\mathbf{B}, \: \lambda > 0$.
Then
\begin{eqnarray}
\ln\mbox{det}_{ren} & \underset{\lambda>>1}\leq & \frac{\lambda^2T\| \mathbf{B}
\|^2}{24\pi^2} \ln \left( \frac{4\lambda^4 \| \mathbf{B} \|^4}{m^2}
\right) \nonumber \\
&    & +\frac{T}{\pi} \int^{\infty}_{\lambda^4 \| \mathbf{B} \|^4 }
\frac{dM^2}{\sqrt{M^2-m^2}} \ln\mbox{det}_4 (\lambda \mathbf{B},M^2) +
O(\lambda^{-2}).
\nonumber \\
\end{eqnarray}
Evidently the large mass behavior of $\ln\mbox{det}_4$ is needed in (3.13).
Eqs.\ (3.4) and (3.6) can be combined to give
\begin{eqnarray}
\ln\mbox{det}_4(m^2) & = & \frac{1}{2}\int_0^\infty \frac{dt}{t} 
\left[\vphantom{\frac{d^{3^3}k}{(2\pi^3_1}} \mbox{Tr}\left(e^{-P^2t}
-exp\{-[(\mathbf{P}-\mathbf{A})^2-\boldsymbol{\sigma}\cdot\mathbf{B}]t\} \right)
\right. \nonumber \\
&   & + \left. \frac{t^{1/2}}{2\pi^{3/2}} \int^1_0 dz z(1-z) \int \frac
{d^3k}{(2\pi)^3} |\mathbf{\hat{B}(k)} |^2 \:
e^{-k^2 \, z(1-z)t} \right] e^{-tm^2},
\nonumber \\
\end{eqnarray}
so that the large mass limit will come from the small-$t$ region of
$\mbox{det}_4\mbox{'s}$ proper time representation.  Carrying out the small-$t$
expansion we find
\begin{eqnarray}
\mbox{Tr} \left( \exp \left\{- [(\mathbf{P}-\mathbf{A})^2-
\boldsymbol{\sigma}\cdot\mathbf{B}]t \right\} -e^{-P^2t} \right) 
\nonumber \\
 =  (4\pi t)^{-3/2} \int d^3r \left[ \frac{2}{3} t^2
\mathbf{B}\cdot\mathbf{B} + \frac{2}{15} t^3
\mathbf{B}\cdot\nabla^2\mathbf{B} \right.  \nonumber \\
   -\frac{2}{45} t^4 (\mathbf{B}\cdot\mathbf{B})^2 + \frac{1}{70} \: t^4
\mathbf{B}\cdot\nabla^4\mathbf{B} &  & \nonumber \\
\left.\vphantom{\frac{2}{3}} +O(t^5\mathbf{B}\cdot\mathbf{B} \, 
\mathbf{B}\cdot\nabla^2
\mathbf{B}, \: t^5 \mathbf{B}\cdot\nabla^6 \mathbf{B}) \right], 
\end{eqnarray}
which, together with (3.14), gives the large-mass expansion of
$\ln\mbox{det}_4$:
\begin{eqnarray}
\ln\mbox{det}_4 & = & \frac{1}{2} \int^\infty_0 \frac{dt}{t} \int d^3r \left[
\frac{t^{5/2}}{180\pi^{3/2}} (\mathbf{B}\cdot\mathbf{B})^2 +O(t^{7/2}
\mathbf{B}\cdot\mathbf{B}\:\mathbf{B}\cdot\nabla^2\mathbf{B})\right]
e^{-tm^2} \nonumber \\
& = & \frac{\int(\mathbf{B}\cdot\mathbf{B})^2}{480\pi \, |m|^5}
+O \left( \frac{\int\mathbf{B}\cdot\mathbf{B} 
\mathbf{B}\cdot\nabla^2\mathbf{B}}{|m|^7} \right).
\end{eqnarray}
Then
\begin{eqnarray}
\int^\infty_{ \lambda^4 \| \mathbf{B} \| ^4} \frac{dM^2}{\sqrt{M^2-m^2}} 
\ln\mbox{det}_4 (\lambda\mathbf{B},M^2) = 
\frac{\int (\mathbf{B}\cdot\mathbf{B})^2}{960\pi \|
\mathbf{B} \|^8 \lambda^4} + O(\lambda^{-8}),
\end{eqnarray}
and hence (3.13) and (3.17) give the bound in Table I:
\begin{equation}
\lim_{\lambda\rightarrow\infty}
\frac{\ln\mbox{det}_{ren}(\lambda\mathbf{B})}{\lambda^2\ln\lambda} \leq
\frac{T \| \mathbf{B} \|^2}{6\pi^2}.
\end{equation}
We note that this upper bound for a general static field $\mathbf{B}$
is greater by a factor of two than the case when $\mathbf{B}$ is 
unidirectional [1].

Let us conclude this section with a comment on the physics of (3.18). 
The main input was the ``diamagnetic'' bound given by the upper bound 
in (3.10).  It is a reflection of the paramagnetic tendency of charged
fermions in an external magnetic field as evident from (3.4):  the
eigenvalues of the Pauli Hamiltonian are, on average, reduced in the
presence of $\mathbf{B}$ relative to the $\mathbf{B}=0$ case.  The
bound in (3.18) is saying that because of this there is a limit  on
how fast the one-loop effective action---due to the vacuum fermion
current density induced by $\mathbf{B}$---can grow.  It is also interesting
that the diamagnetic bound has come to us by a long chain of reasoning
starting with QED$_3$ on a lattice, that it had lain dormant for about
seventeen years, and then emerged again to tell us something nontrivial
about QED$_4$.

\subsection{Zeros of \mbox{det}$\mathbf{_4}$}

In order to write (3.9) in the form (3.10) it is necessary to show that
$\mbox{det}_{QED3}$ or, equivalently $\mbox{det}_4(1-eS\hspace{-0.5em}\Asla)$,
has no zeros for real $e$ when $m \neq 0$. 
Instead of working in the Hilbert space 
$L^2(\R^3, \: \sqrt{k^2+m^2} \: d^3k; \C^2)$ introduced in Section 3.2 we will
make a similarity transformation on $S\hspace{-0.5em}\Asla$, which does not 
change its eigenvalues, and deal with the integral operator 
\begin{equation}
K=(P^2+m^2)^{1/4} S\hspace{-0.5em}\Asla(P^2+m^2)^{-1/4},
\end{equation}
on $L^2(\R^3,d^3r;\C^2)$ [21].  Let $\psi_n \in L^2$ be an eigenvector
of $K$,
\begin{equation}
K\psi_n = \frac{1}{e_n}\psi_n.
\end{equation}
Taking the Fourier transform of (3.20) gives
\begin{equation}
\int \frac{d^3k}{(2\pi)^3} \hat{\hspace{-0.5em}\Asla}
(\mathbf{p}-\mathbf{k}) (k^2+m^2)^{-1/4}
\hat{\psi}_n (\mathbf{k}) = \frac{1}{e_n} \frac{\psl+m}{(p^2+m^2)^{1/4}}
\hat{\psi}_n(\mathbf{p}).
\end{equation}
Its complex conjugate is
\begin{equation}
- \int \frac{d^3k}{(2\pi)^3} \hat{\psi}^\dagger_n (\mathbf{k}) (k^2+m^2)^{-1/4}
\hat{\hspace{-0.5em}\Asla} (\mathbf{k}-\mathbf{p}) 
= \frac{1}{e^\star_n} 
\frac{\hat{\psi}^\dagger_n (\mathbf{p}) (m-\psl)}{(p^2+m^2)^{1/4}}.
\end{equation}
Multiply (3.21) from the left by $\hat{\psi}^\dagger
(\mathbf{p})(p^2+m^2)^{-1/4}$ and (3.22) from the right by 
$(p^2+m^2)^{-1/4} \hat{\psi}_n (\mathbf{p})$; add the two equations and
integrate both sides over $p$ to get
\begin{eqnarray}
i\mbox{Im} \, (e_n) \int d^3p(p^2+m^2)^{-1/2} \hat{\psi}_n^{\dagger} 
(\mathbf{p})\psl
\hat{\psi}_n (\mathbf{p}) \nonumber \\
\ \ \ \ \ \ = m \mbox{Re} \, (e_n) \int d^3p \: (p^2+m^2)^{-1/2} 
| \hat{\psi}_n (\mathbf{p}) |^2.
\end{eqnarray}
Since $\psi_n \in L^2$ so does $\hat{\psi}_n$.  Therefore both integrals
in (3.23) converge by inspection, and the integral on the right-hand
side is not zero.  Since $\overset{\infty}{\underset{n=1}{\sum}} |e_n|^{-p}
< \infty$ for
$p > 3$, there are no zeros at the origin.  Hence (3.23) implies 
$\mbox{Im}(e_n) \neq 0$ if $m \neq 0$.  A similar conclusion was reached in
QED$_4$ by Alder [26].

In Section 3.2 we saw that, for the potentials we are considering,
$S\hspace{-0.5em}\Asla \in \mathcal{C}_p, \  p >3.$
Then by Theorem 6.2 of Simon in
ref.\ [24] we can express $\mbox{det}_4$ in terms of the eigenvalues of 
$S\hspace{-0.5em}\Asla$:
\begin{eqnarray}
\mbox{det}_4 (1-eS\hspace{-0.5em}\Asla) = \prod^\infty_{n=1} 
\left[ \left( 1 - \frac{e}{e_n} \right) exp
\left(\sum^3_{k=1} (e/e_n)^k/k \right) \right].
\end{eqnarray}
Since all of the eigenvalues are off the real axis for $m \neq 0$,
det$_4$ cannot vanish for real values of $e$, and therefore (3.10)
follows from this, definition (3.6) and det$_{QED3}|_{e=0}=1$.

Since there is a nonsingular matrix $C$ such that 
$C^{-1}\gamma_\mu C=-\gamma_\mu^T$, namely $C=\gamma_2$ in the representation 
$\gamma_\mu=(i\sigma_1, i\sigma_2, i\sigma_3), \ C\mbox{-invariance}$
is maintained and hence det$_4$ is an even function of $e$.  This and
the reality of det$_4$ for real $e$ imply that the eigenvalues appear in
quartets $\pm e_n, \ \pm e_n^\star$.  The same conclusion in
QED$_4$ was reached by the authors in ref.\ [27].

It is not essential that $\mathbf{A} \in L^3(\R^3)$.  Instead, one may
just assume that $\mathbf{A} \in L^6(\R^3)$ in the Coulomb gauge as
required if $\mathbf{B}$ is square integrable.  In this case
$S\hspace{-0.5em}\Asla \in \mathcal{C}_6$
so that one must deal with det$_6$, whose expansion
begins in sixth order.  The analysis above is trivially modified with
the end result that (3.18) is unchanged.

Finally, the analysis we have used to show that det$_{QED3}$ has no
zeros for real $e$ and $m \neq 0$ may be applied to det$_{QED2}$.  Here
it must be kept in mind that $A_\mu$ behaves as a ``winding'' field in
the gauge $\partial_\mu A^\mu =0$ with a $1/r$ fall off if the magnetic
flux is nonvanishing.  By assuming $A_\mu \in L^3(\R^3)$ one can show
that $S\hspace{-0.5em}\Asla \in \mathcal{C}_3$ and conclude that
det$_{QED2}$ is never negative if det$_{QED2}|_{e=0}=1$.

\section*{References}

\begin{description}
\item [{[1]}] M. P. Fry, Phys.\ Rev.\ D{\bf 45}, 682(1992).
\item [{[2]}] M. P. Fry, Phys.\ Rev.\ D{\bf 47}, 2629(1993).
\item [{[3]}] M. P. Fry, Phys.\ Rev.\ D{\bf 51}, 810(1995).
\item [{[4]}] M. P. Fry, Phys.\ Rev.\ D{\bf 53}, 980(1996).
\item [{[5]}] J. Schwinger, Phys.\ Rev.\ {\bf 82}, 664(1951).
\item [{[6]}] E. Seiler, in {\it Gauge Theories:  Fundamental
Interactions and Rigorous Results}, Proceedings of the International
School of Theoretical Physics, Poiana Brasov, Romania, 1981, edited by
P. Dita, V. Georgescu, and P. Purice, Progress in Physics Vol.\ 5
(Birkh\"{a}user, Boston, 1982), p.263.
\item [{[7]}] W. Dittrich and M. Reuter, {\it Effective Lagrangians in
Quantum Electrodynamics},  Lecture Notes in Physics Vol.\ 220 (Springer,
Berlin, 1985).
\item [{[8]}]  D. Bridges, J. Fr\"{o}hlich, and E. Seiler, Ann.\ Phys.\
(N.Y.) {\bf 121}, 227(1979).
\item [{[9]}] J. Schwinger, Phys.\ Rev.\ {\bf 128}, 2425(1962).
\item [{[10]}] J.E. Avron and R. Seiler, Phys.\ Rev.\ Lett.\ {\bf 42},
931(1979).
\item [{[11]}] K. Miller, ``Bound States of Quantum Mechanical Particles
in Magnetic Fields'', Ph.D. Thesis, Princeton University (1982).
\item [{[12]}] D. Cangemi, E. D'Hoker, and G. Dunne, Phys.\ Rev.\
D{\bf 52}, R3163(1995).
\item [{[13]}] D. H. Weingarten, Ann.\ Phys.\ (N.Y.) {\bf 126},
154(1980).
\item [{[14]}] H. Euler and B. Kockel, Naturwiss. {\bf 23}, 246(1935); W.
Heisenberg and H. Euler, Z.\ Phys.\ {\bf 98}, 714(1936); V. Weisskopf, K.
Dan.\ Vidensk.  Selsk.\ Mat.-fys.\ Medd.\ {\bf 14}, No.\ 6(1936).
\item [{[15]}] S. Deser, R. Jackiw, and S. Templeton, Ann.\ Phys.\
(N.Y.) {\bf 140}, 372(1982).
\item [{[16]}] A.N. Redlich, Phys.\ Rev.\ D{\bf 29}, 2366(1984).
\item [{[17]}] R. R. Parwani, Phys.\ Lett.\ B{\bf 358}, 101(1995).
\item [{[18]}] G. Dunne and T. Hall, ``Inhomogeneous Condensates in
Planar QED'', U. Conn.\ preprint, hep-th/9511192.
\item [{[19]}] E. H. Lieb and W. E. Thirring, in {\it Studies in
Mathematical Physics, Essays in Honor of Valentine Bargmann}, edited by E. H.
Lieb, B. Simon, and A. S. Wightman (Princeton U.\ Press, Princeton,
1976), p.\ 269.
\item [{[20]}] M. Reed and B. Simon, {\it Methods of Modern Mathematical
Physics II:  Fourier Analysis, Self-Adjointness}  (Academic Press, Inc.\,
New York, 1975)
\item [{[21]}] E. Seiler and B. Simon, Commun.\ Math.\ Phys.\
{\bf 45}, 99(1975).
\item [{[22]}] E. Seiler, {\it Gauge Theories as a Problem of
Constructive Quantum Field Theory and Statistical Mechanics}, Lecture
Notes in Physics Vol.\ 159 (Springer, Berlin/Heidelberg/New York,
1982).
\item [{[23]}] B. Simon, {\it Trace Ideals and their Applications},
London Mathematical Society Lecture Note Series 35 (Cambridge U. Press,
Cambridge, 1979).
\item [{[24]}] B. Simon, Advances in Math.\ {\bf 24}, 244(1977).
\item [{[25]}] E. Seiler, private communication.
\item [{[26]}] S. L. Adler, Phys.\ Rev.\ D{\bf 16}, 2943(1977).
\item [{[27]}] C. Itzykson, G. Parisi, and J.-B. Zuber, Phys.\ Rev.\
D{\bf 16}, 996(1977); R. Balian, C. Itzykson, J.-B. Zuber, and G. Parisi,
{\it ibid.} {\bf 17}, 1041(1978).
\end{description}

\begin{table}
\begin{center}
\begin{tabular}{ll}
\hline
$\text{QED}_2$
& 
\begin{minipage}{4in}
\begin{equation*}
-\frac{\|B\|^2}{4 \pi m^2} \leq \ln \det  \leq 0 
\end{equation*}
\end{minipage}
\\
$\text{QED}_3$
&
\begin{minipage}{4in}
\begin{equation*}
- \frac{Z}{6 \pi } \int d^2r |B|^{3/2} \leq \ln \det \leq 0  
\end{equation*}
\end{minipage}
\\
$\text{QED}_4$
&
\begin{minipage}{4in}
\begin{equation*}
\frac{ZT\|B\|^2}{48 \pi^2} \leq \lim_{\lambda \rightarrow \infty} 
\left( \frac{\ln \mbox{det}_{ren}(\lambda B)}{\lambda^2 \ln \lambda} \right)
\leq \frac{T\| \bfB \|^2}{6 \pi^2} 
\end{equation*}
\end{minipage}
\\
\hline
\end{tabular}
\end{center}
\caption{
Bounds on fermionic determinants.  The lower bounds in 
QED$_2$ (Ref.\ [4]), QED$_3$ (see Section 2) and QED$_4$ (Ref.\ [4]) are for the field 
$\bfB =(0,0,B(x,y)), \ B(x,y) \geq 0$.  For $B(x,y) \leq 0$
replace $B$ with $-B$.  The upper bound for QED$_2$ (Refs. [8,13]) has no 
restriction on the sign of
$B(x,y)$.  The upper bounds for QED$_3$ (Ref.\ [8]) 
and QED$_4$ (see Section 3) are for a static, directionally-varying
field $\bfB ( \bfr )$, $ \bfr \in R^3 $.  $Z$ and $T$ denote the 
size of the boxes
in the z-and t-directions.  
The lower bounds for QED$_{2,3}$ are representative; 
better but more complicated bounds may be found in Section 2 and in Ref.\ [4].}
\end{table}

\end{document}